\newcommand{\comment}[1]{}

\newcommand{\myshowfig}[1]{}

\def\qrmsps{Q_{rms-PS}}

\def\bis{$I^3_{\l}$}
\def\bisxvi{$I^3_{16}$}

\def\etal{{\frenchspacing\it et al.}}

\def\eg{{\frenchspacing\it e.g.}}

\def\beq#1{\begin{equation}\label{#1}}
\def\eeq{\end{equation}}
\def\beqa#1{\begin{eqnarray}\label{#1}}
\def\eeqa{\end{eqnarray}}
\def\eq#1{equation~(\ref{#1})}

\def\sec#1{Section~\ref{#1}}

\def\expec#1{\langle#1\rangle}
\def\l{\ell}
\def\pmax{p_{\rm max}}
\def\x{{\bf x}}
\def\N{{\bf N}}
\def\Smtrx{{\bf S}}
\def\C{{\bf C}}

\def\rn{\noindent\parshape 2 0truecm 8.8truecm 0.3truecm 8.5truecm}
\def\nn#1 #2{#1, #2.}				
\def\nnn#1 #2 #3{#1, #2. #3.}			
\def\nnnn#1 #2 #3 #4{#1, #2. #3. #4.}		
\def\nnnnn#1 #2 #3 #4 #5{#1, #2. #3. #4. #5.}	
\def\dualand{, \&\hbox{ }}				
\def\multiand{, \&\hbox{ }}				

\def\rg#1;#2;#3;#4;#5;#6 {\par\rn#1 #2, {\it #3}, {\bf #4}, #5 (``#6'') \par}
\def\rf#1;#2;#3;#4;#5 {\par\rn#1 #2, {\it #3}, {\bf #4}, #5\par}
\def\rfbook#1;#2;#3;#4;#5 {{\frenchspacing\par\rn#1 #2, {\it #3} (#4: #5)\par}}
\def\rfproc#1;#2;#3;#4;#5;#6 {{\frenchspacing\par\rn#1 #2, in {\it #3}, ed. #4 (#5: #6)\par}}
\def\rfprep#1;#2;#3  {{\par\rn#1 #2, #3\par}}
\def\rfprepp#1;#2;#3 {{\par\rn#1 #2, #3\par}}

\documentstyle[emulateapj,danonecolfloat]{article}
\begin{document}
\twocolumn[

\title{
   Is the cosmic microwave background really non-Gaussian?
}

\author{
Benjamin C. Bromley\footnote{Department of Physics, University of Utah,
201 JFB, Salt Lake City, UT 84112; bromley@physics.utah.edu}
and
Max Tegmark\footnote{Institute for Advanced Study, Princeton, 
NJ 08540; max@ias.edu}$^,$\footnote{Hubble Fellow}
}


\submitted{Submitted to ApJL April 20, 1999; accepted June 28}

\begin{abstract}
Two recent papers have claimed detection of non-Gaussian features in
the COBE DMR sky maps of the cosmic microwave background. We confirm
these results, but argue that Gaussianity is still not convincingly
ruled out.  Since a score of non-Gaussianity tests have now been
published, one might expect some mildly significant results even by
chance.  Moreover, in the case of one measure which yields a
detection, a bispectrum statistic, we find that if the non-Gaussian
feature is real, it may well be due to detector noise rather than a
non-Gaussian sky signal, since a signal-to-noise analysis localizes it
to angular scales smaller than the beam.  We study its spatial origin
in case it is nonetheless due to a sky signal (\eg, a cosmic string
wake or flat-spectrum foreground contaminant). It appears highly
localized in the direction $b = 39.5^\circ$, $l = 257^\circ$, since
removing a mere 5 pixels inside a single COBE beam area centered there
makes the effect statistically insignificant. We also test Guassianity
with an eigenmode analysis which allows a sky map to be treated as a
random number generator. A battery of tests of this generator all
yield results consistent with Gaussianity.
\end{abstract}

]

\section{Introduction}
\label{sec:intro}

The detection of fluctuations in the cosmic microwave background (CMB)
by the Differential Microwave Radiometer (DMR) on board the Cosmic
Microwave Explorer (COBE) satellite (Smoot \etal\ 1992) began a new
era for studies of the Universe on large scales. Current and planned
experiments offer the promise of tight constraints on cosmological
parameters, including both quantities important for observational
astronomy and inflationary parameters which may provide the ultimate
testing ground for fundamental particle physics, relating phenomena on
the largest and smallest observable scales.

A key cosmological constraint from observations of the primordial
density field is its general statistical nature. In most inflationary
scenarios, the density is a Gaussian random field, (although see, \eg,
Peebles 1999).  This implies that the joint probability distribution
of the $N\sim 4000$ temperatures in the galaxy-cut DMR sky map is a
multivariate Gaussian.  In contrast, topological defect models predict
a non-Gaussian density field (\eg, Avelino {\etal} 1998). However, an
analysis by Pen {\etal} (1997) suggests that defects are inconsistent
with the observed power spectrum of fluctuations. At present,
inflationary models, or at least generically Gaussian models, dominate
the literature on large-scale structure.

It is therefore quite intriguing that two groups, Ferreira, Magueijo
\& Gorski (1998, hereafter FMG) and Pando, Valls-Gabaud \& Fang (1998,
hereafter PVF), claim that the CMB fluctuations measured by COBE DMR
are non-Gaussian. If substantiated, these results could potentially
rule out standard inflation as the primary mechanism for cosmic
structure formation in the early Universe.  The results are all the
more surprising because the literature documents considerable previous
effort to identify non-Gaussianity in the CMB (see references given in
\S\ref{DebunkSec}), all of which failed.

In this {\em Letter}, we revisit the question of whether COBE DMR
rules out Gaussianity.  We begin with a series of new tests of
Gaussianity based on eigenmode analyses of the DMR map
(\S\ref{EigenSec}).  We then consider the results of FMG and PVF in
\sec{DebunkSec}, exploring their sensitivity to individual modes and
spatial features as well as the significance of these results in light
of the other statistical tests which have been applied to the
data.  We present our conclusions in \sec{ConclusionSec}.

\section{Statistics of eigenmodes}\label{EigenSec}

We analyze a sky map formed of the combined 53 and 90 GHz four year DMR
data pixelized at resolution 6 in Galactic coordinates, with the
``custom'' Galaxy sky mask (Bennett {\etal} 1996), and with monopole
and dipole contributions removed as in Tegmark \& Bunn (1995).  The
resulting data set consists of temperatures for 3,881 pixels in the
sky, which we arrange in a vector $\x$.  We take the covariance matrix
$\C\equiv \expec{\x\x^t}$ to be of the form $\C=\Smtrx+\N$, where the
noise matrix $\N$ is diagonal and the signal matrix $\Smtrx$ is a
Harrison-Zel'dovich power spectrum normalized to $\qrmsps = 18 \ \mu$k
(\eg, Bennett {\etal} 1996).

The pixelized data are both correlated and noisy, hence we subject the
map $\x$ to both a principal component analysis (PCA) and a
signal-to-noise eigenmode analysis (SNA), two standard astrophysical
tools.  Both of these procedures involve expanding the data in a new
basis, the eigenvectors of $\C$ for the PCA case and the eigenvectors
of $\N^{-1/2}\Smtrx\N^{-1/2}$ for the SNA case.  The eigenvectors are
sorted by decreasing eigenvalue and normalized so that the expansion
coefficients have unit variance.  For the SNA (Bond 1994; Bunn \&
Sugiyama 1995; Tegmark {\etal} 1997), the modes are listed in order of
decreasing signal-to-noise level.  For the PCA, the modes explain
successively less and less of the variance in the data. Since the DMR
noise per pixel does not fluctuate much, the two methods give similar
results.  The first few hundred modes contain essentially all the
cosmological information, and probe successively smaller angular
scales (Bond 1994; Bunn \& Sugiyama 1995; Bunn \& White 1996).  We use
the top 250 modes for the Gaussianity tests described below.

The purpose of this exercise is two-fold: First, we can determine how
many cosmologically significant degrees of freedom a given statistical
test should consider.  Second, the decomposition into uncorrelated
eigenmodes allows the data to be cast as a list of random numbers
which, under the null hypothesis that the DMR data are
Gaussian\footnote{Note that as long as the true CMB sky is Gaussian,
our data set will be Gaussian as well: Both the smoothing done by the
DMR beam, our galaxy cut and our monopole and dipole removal are
linear operations, and all linear operations preserve Gaussianity.},
will be independent and normally distributed.  Although
the statistics of these samples may not be completely testable in
practice, we can still constrain general properties of the COBE data.

We run both the lists of 250 PCA and SNA entries and the entire list
of 3877 numbers (the rank of the covariance matrix after monopole and
dipole subtraction) from the PCA basis through a smattering of tests,
first for Gaussianity of the individual list elements. The null
hypothesis cleanly passes Kolmogorov-Smirnoff and $\chi$-square tests,
along with tests of cumulants up to fourth order and of the
significance of the top few outliers.  None of these tests manage to
reject the Gaussian null hypothesis with 95\% confidence.  Note that
these tests are sensitive only to the 1-point distribution of mode
amplitudes, not to correlations between modes.  This is strong though
not irrefutable evidence that if the DMR data are non-Gaussian then
mode correlations, not mode amplitudes, are responsible.

The next step in testing the Gaussian hypothesis is to look for mode
correlations. This is a difficult thing to do in any exhaustive way,
even for our short lists of 250 elements, and a thorough treatment of
this problem is beyond the scope of this {\em Letter}.  We note only
that no correlations were detected above the 95\% confidence level in
tests of second, third and fourth order $N$-point correlations.  We
also used mode amplitudes to simulate rolls of a die and examine the
one- and two-point distributions of outcomes to see if the die is
loaded.  If one wishes to play dice with the Universe, evidently it
would be a fair game.

\begin{figure*}[tb] 
\centerline{\epsfxsize=12cm\epsffile{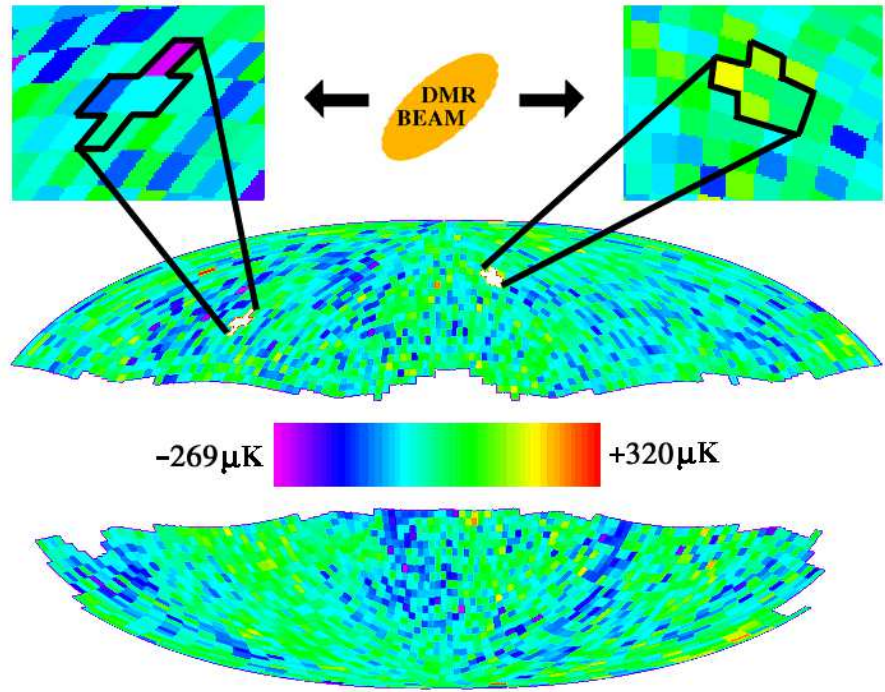}}
\figcaption{
The galaxy-cut COBE DMR map is shown in Aitoff projection.
The inset shows the five pixels whose omission makes the 
FMG effect go away, with the ellipse indicating the 
FWHM size and the projected shape of the DMR beam
in that region.
}
\end{figure*}

\section{Reports of Non-Gaussianity}
\label{DebunkSec}

The above tests of the COBE data (26 in all) join a sizable list of
previous results without significant detections of non-Gaussianity.
For the 53 GHz DMR 1 year data, the three-point function was studied
by Luo (1994, one test) and Hinshaw {\etal} (1994, two tests) while
Smoot {\etal} (1994) considered the topological genus and kurtosis
(two tests).  For the 53 and 90 GHz DMR 2 year data, Hinshaw {\etal}
(1995) studied the equilateral and pseudocollapsed three-point
function at three $\l$-cuts with 12 tests in total; the most extreme
gave a 98\% non-Gaussianity detection, but was deemed to suffer from a
known noise problem.  Kogut \etal\ (1996) tested the DMR 4 year data
for three-point correlations, genus and peak correlations (4 tests in
all), while Heavens (1998) analyzed this same data set using an
optimized bispectrum statistic on 5 different scales.  Gazta\~naga
{\etal} (1998) performed 5 variance-of-variance tests with the
strongest rejection of Gaussianity being at the 91\% level. Most
recently, Diego {\etal} (1999) concluded that the DMR data were
consistent with Gaussianity in a partition function analysis.

On the other hand, there are two Gaussianity tests which the data
reportedly fail. One is based on the bispectrum statistic of FMG (at
98\% confidence) and other on the fourth-order wavelet statistic
proposed by PVF (at 99\% confidence).  These are strong signals of
non-Gaussianity, but some caution is in order.  Given that the DMR
data have been subjected to the 32 other published tests cited above
which provide no evidence of non-Gaussianity (not to mention the 26
reported here and any tests, including some of our own, which were not
reported because they yielded null results), are the FMG and PVF
results simply expected outliers in the distribution of test results?

To address this point, suppose we try to rule out some null hypothesis
by subjecting a data set to $n$ independent statistical tests, and
that the most successful one rules it out at a confidence level $p$,
say 99\%.  How significant is this really?  Let $p_i$ denote the
confidence level obtained from the $i^{th}$ test, with
$\pmax\equiv\max\{p_i\}$ corresponding to the most successful test.
The probability of getting a less extreme result is then
\beq{DilutionEq}
P(\pmax<p) = \prod_{i=1}^n P(p_i<p)
= \prod_{i=1}^n p
= p^n \, .
\eeq
For example, the most extreme S/N eigenmode coefficient in
\sec{EigenSec} is a 3.3-$\sigma$ outlier.  If that one coefficient was
all we had, then we would reject Gaussianity at the 99.9\% level.
However, we have 250 independent numbers and \eq{DilutionEq} shows
that our level of confidence in rejection from that one extreme
coefficient is only $0.999^{250}\approx 78\%$.  Similarly, the list of
34 published Gaussianity tests mentioned above contains one which
rules out the null hypothesis at the 99\% level. If these tests were
independent, we could reject Gaussianity with only $0.99^{34}\approx
71\%$ confidence.  Of course the tests are not strictly independent.
Yet if some of them capture only subsets of the information contained
in the $\sim 4000$ COBE data points, then they may be effectively
independent of each other. With this in mind we examine the FMG and
PVF tests in more detail.

\subsection{The wavelet test}

The detection reported by PVF of a non-Gaussian signal is made with a
measure of fourth-order correlations between wavelet coefficients of
the DMR data. The coefficients are obtained using a discrete transform
of the northern sixth of the sky map, after the spherical plane of the
sky has been projected onto the face of a cube.  Taken alone, this
measure reportedly gives a detection of non-Gaussian signal at 99\%
confidence. But we have more information, even about PVF's wavelets: a
second projection of the sky map onto the opposite face of the cube
(i.e., the opposite hemisphere of the sky) lies roughly at the 40\%
confidence level. Together, a joint two-faced wavelet analysis gives a
weaker rejection of the Gaussian hypothesis, formally at 97\%
confidence. Furthermore, the PVF detection is claimed only for
wavelets on one specific scale even though they seek similar
detections on two other scales but do not find them. Likewise, a
detection was sought but not found for third-order moments.  With
$n=2\times 3\times 2$, \eq{DilutionEq} predicts a much lower
confidence level, $0.99^{12}\approx 89\%$, for the claimed detection.

In an analysis based on compact but smooth wavelets (e.g., Bromley
1994), we confirm the existence of a strong (99.6\%) non-Gaussian
outlier at the 11$^\circ$+22$^\circ$ scales reported by PVF.
Interestingly, we can make the entire non-Gaussian signal vanish by
simply zeroing or flipping the sign of a single, modestly rare
principal component amplitude (a 2.7-$\sigma$ fluctuation of the
90$^{\rm th}$ mode).  Also the 17$^{\rm th}$ eigenmodes from both the
PCA and SNA strongly affect the wavelet detection. In the S/N basis,
this mode has an amplitude of 2.1-$\sigma$ (slightly less in the
principal component vector); by zeroing the amplitude of these modes,
the wavelet statistic yields less than a 2-$\sigma$ detection.
Furthermore, when set to zero, a single spot in the sky the size of
the COBE beam (centered at $b = 64.6^\circ$, $l = 40.6^\circ$), also
cuts the non-Gaussian signal down to a similar level.  Of the six
pixels in this spot, three are within one standard deviation of the
expected noise, while two are at the 2.6-$\sigma_n$ level and the
third is a strong outlier at 3.1-$\sigma_n$.  (How this spot affects
the PVF results depends on unspecified details of their analysis.)

\subsection{The bispectrum test}

FMG introduce a measure, \bis, based on averaged triplets of
projection coefficients from even-multipole spherical harmonics.
Non-Gaussian behavior is seen only at $\l = 16$, but FMG are careful
to consider the fact that the bispectrum at eight other $\l$-values
are individually consistent with the Gaussian hypothesis. The reported
confidence of the non-Gaussian detection is 98\%.

There is nonetheless a possibility that the nine \bis\ values given by
FMG are sensitive to only a fraction of the information in the COBE
data. Although there is some spherical harmonic mode coupling as a
result of the sky mask, the odd $\l$ multipoles are largely missing as
well as multipoles above $\l = 18$.  There may also be dependence on
localized noise. We emphasize this latter point by setting to zero the
five pixels in the beam-size spot on the sky centered at Galactic
latitude $b = 39.5^\circ$ and longitude $\l = 257^\circ$, shown in
Fig.~1.  The value of \bisxvi\ falls from about 0.92 to 0.78,
approximately a 98\% detection on its own\footnote{Note that the
effect of zeroing the spot depends somewhat on details of removing
monopole and dipole contributions to the DMR maps.  If explicit
removal (e.g., Tegmark \& Bunn 1995) is not performed on the custom
cut map, as is apparently the case in the bispectrum analysis of
Magueijo, Ferreira \& Gorski (1999; Fig.~1 therein), then the effect
of removing the spot (actually a nearest-neighbor) is to lower
\bisxvi\ to 0.66.}, but well below 2-$\sigma$ when taken in
conjunction with the other eight \bis\ values shown in Fig.~2
($.98^9\approx 84\%$ if the 9 values were uncorrelated).  Note that
there is a single, rare pixel brightness value in the spot.  In
units of the expected noise fluctuations, it is at the level of
$-3.5\sigma_n$, and zeroing it alone cuts \bisxvi\ by 10\%.

\begin{figure}[tb]
\vskip-2.7cm
\hglue-0.8cm
\centerline{\hglue1cm{\vbox{\epsfxsize=10.5cm\epsfbox{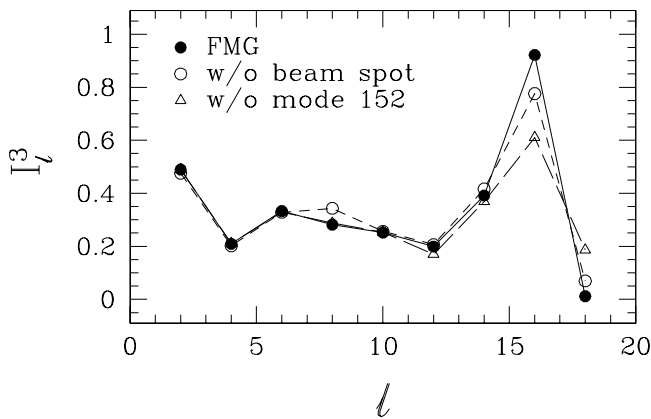}}}}
\vskip-2.3cm
\figcaption{
The bispectrum statistic of FMG for various multipoles $\l$.
The dark circles are the full galaxy-cut COBE data (cf. Fig.~1 in FMG), and 
the open circles are the data after zeroing 5 pixels inside a single
COBE beam centered at $b = 39.5^\circ$, $l = 257^\circ$). The
triangles result when the amplitude of mode 151 (containing 0.15\% of
the total power) is set to zero.
}
\label{fig:bis}
\end{figure}

The bispectrum statistic is also highly sensitive to individual
eigenmodes.  Zeroing or flipping the sign of principal component 151
causes \bisxvi\ to drop from 0.92 to the unambiguously Gaussian values
of 0.61 and 0.20, respectively.
Furthermore, there is sensitivity to S/N eigenmode 224, a 3-$\sigma$
fluctuation (the second most extreme of the first 250 modes) which is
more strongly coupled to noise than cosmic structure (its S/N
eigenvalue is 0.4). Zeroing this knocks \bisxvi\ to 0.74; flipping the
sign causes the value to fall to 0.52, a decidedly Gaussian value.
Since both of these eigenmodes are dominated by noise rather than
cosmic signal, it is possible that the source of the alleged
non-Gaussianity is detector noise rather than CMB.

Note that in the above analyses, we systematically searched for the
modes or pixels which affected the wavelet and bispectrum statistics
the most.  In each case we found only a few which cause more than an
insignificant (several percent) change in these measures, and
interestingly the significant modes or pixels were different for the
two measures.  
Using randomly generated Gaussian skymaps selected for
apparent non-Gaussianity similar to the DMR data, we also checked that
it is quite common for a Gaussian map to give a wavelet or bispectrum
measure that is sensitive to only a few individual pixels or modes,
just as in the DMR case.

\section{Conclusion}
\label{ConclusionSec}

The problem of the statistical nature of the CMB may be cast in a
conceptually simple form: just use the observed temperature
fluctuations to make a random number generator, based on assumed
statistical properties, and test its quality.  Here we have used the
53+90 GHz COBE sky map to generate lists of putative random numbers
with principal components and signal-to-noise eigenmodes.  In both
cases the one-point distribution is manifestly Gaussian.  If the CMB
on COBE scales is non-Gaussian, it is the result of correlations
between modes.  Unfortunately, exhaustive tests of mode correlations
are not feasible. A few tests which can pick up a fair range of
non-Gaussian behavior were performed and no evidence of mode
correlations was found.

Here we have also considered the two statistics which reportedly detect
non-Gaussianity in the COBE data.  Both detections turn out to be
fragile in the sense that they vanish when a single DMR beam spot or
a single eigenmode is removed.  Moreover, we found that the detection
by PVF, based on wavelets alone, was less significant than originally
claimed.  Even so, with the dozens of different Gaussianity tests that
have now been published, it would not be surprising if a perfectly
valid analysis rejected Gaussianity at say 98\% confidence purely by
accident.

Our results cast some doubt on the significance of the claimed
non-Gaussian behavior in the CMB.  If the reported detections are real
nonetheless, then the eigenmodes and COBE-beam spots that we isolated
for the wavelet and bispectrum statistics are candidates for potential
non-Gaussian sources in the CMB. This latter possibility would perhaps
be more satisfying if both measures were coupling to the same
non-Gaussian structure in the sky. However, this is not obviously the
case, since both the bispectrum statistic and the wavelet measure show
virtually no sensitivity to the sky spots and eigenmodes which so
dramatically affect the other.

It is generally much easier to show that a bad random number generator
is bad then to prove that a good one is good. Indeed, the results
reported here fail to demonstrate that the CMB really is
Gaussian. Conversely, the search for non-Gaussianity is also something
of an uphill battle, a fight against the central limit theorem which
causes both instrumental effects and the linear combinations involved
in the eigenmode expansions to make things look more Gaussian.
Therefore statistical measures should be tuned for the specific type
of non-Gaussianity that physical models predict. This approach is
taken in many recent studies (\eg, Cayon \& Smoot 1995; Magueijo 1995;
Torres {\etal} 1995; Gangui 1996; Gangui \& Mollerach 1996; Ferreira
\& Magueijo 1997; Ferreira {\etal} 1997; Barrieiro {\etal} 1998, Lewin
{\etal} 1999; Popa 1998) with an eye toward upcoming, high-resolution
CMB data.

We thank Ang\'elica de Oliveira-Costa, Al Kogut, Alex Lewin, Bill
Press, George Rybicki, and Nelson Beebe for useful comments.
BCB acknowledges partial support from NSF Grant PHY 95-07695 and the use
of supercomputing resources provided by NASA/JPL and Caltech/CACR.  MT
was funded by NASA though grant NAG5-6034 and Hubble Fellowship
HF-01084.01-96A from STScI, operated by AURA, {\frenchspacing Inc.}
under NASA contract NAS5-26555.
\vskip-1cm



\begin{references}
\bigskip

\rf\nnn Avelino P P, \nnnn Shellard E P S, \nnnn Wu J H P\multiand\nn Allen B;1998;ApJL;507;L101

\rf\nnn Barrieiro R B, \nnn Sanz J L, Mart\'{\i}nez-Gonz\'alez E\multiand\nn Silk J;
1998;MNRAS;296;693


\rf\nnn Bennett C L {\etal};1996;ApJ;464;L1

\rf\nnn Bond J R;1994;Phys. Rev. Lett.;74;4369

\rf\nnn Bromley B C;1994;ApJ;423;L81


\rf\nnn Bunn E F\dualand\nn Sugiyama N;1995;ApJ;446;49

\rf\nnn Bunn E F\dualand\nn White M;1995;ApJ;480;6


\rf\nn Cayon L\dualand\nnn Smoot G F;1995;ApJ;452;487

\rf\nnn Diego J M, \nn Mart\'{\i}nez-Gonz\'alez E,
\nnn Sanz J L, \nn Mollerach S \multiand\nn Mart\'{\i}nez V;1999;MNRAS;306;427

\rf\nnn Ferreira P G\dualand\nn Magueijo J;1997;Phys. Rev. D;55;3358

\rg\nnn Ferreira P G, \nn Magueijo J\multiand\nnn Gorski K M;1998;ApJ;503;1;FMG

\rf\nnn Ferreira P G, \nn Magueijo J\multiand\nn Silk J;1997;Phys. Rev. D;56;4592

\rf\nn Gazta\~naga E, \nn Fosalba P\multiand Elizalde E;1998;MNRAS;295;30P

\rf\nn Gangui A;1996;Helv. Phys. Acta;69;215
 
\rf\nn Gangui A\dualand\nn Mollerach S;1996;Phys. Rev. D;54;4750

\rf\nn Hinshaw G {\etal};1994;ApJ;431;1

\rf\nn Hinshaw G {\etal};1995;ApJ;446;L7

\rf Kogut A {\etal};1995;ApJL;439;29L
		     						     
\rf Kogut A {\etal};1996;ApJL;464;L29
				
\rf\nn Lewin A, \nn Albrecht A\multiand\nn Magueijo J;1999;MNRAS;302;131

\rf\nn Luo X;1994;Phys. Rev. D;49;3810
%

\rfprep\nn Magueijo J, \nnn Ferreira P G\multiand\nnn Gorski K M;1999;astro-ph/9903051

\rf\nn Magueijo J;1995;Phys. Rev. D;52;4361

\rg\nn Pando J, \nnn Valls-Gabaud D\multiand\nn Fang L;1998;
Phys. Rev. Lett.;81;4568;PVF

\rfprep\nn Popa L;1998;astro-ph/9806086

\rf\nnnn Peebles P J E;1999;ApJ;510;523

\rf Pen {U.-L.}, \nn Seljak U\multiand\nn Turok N;1997;Phys. Rev. Lett.;79;1611

\rf\nnn Smoot G F {\etal};1992;ApJL;396;L1

\rf\nnn Smoot G F {\etal};1994;ApJ;437;1

\rf\nn Tegmark M\dualand\nnn Bunn E F;1995;ApJ;455;1

\rf\nn Tegmark M, \nnn Taylor A N\multiand\nnn Heavens A F;1997;ApJ;480;22

\rf\nn Torres S {\etal};1995;MNRAS;274;853       



\end{references}
\end{document}